\documentclass[twocolumn,letter]{jpsj2} 
\title{Low-lying optical phonon modes in the filled skutterudite CeRu$_4$Sb$_{12}$}

\author{C. H. \textsc{Lee}$^{1}$\thanks{E-mail address: c.lee@aist.go.jp}, I. \textsc{Hase}$^{1}$, H. \textsc{Sugawara}$^{2}$, H. \textsc{Yoshizawa}$^{3}$, and H. \textsc{Sato}$^{4}$}

\inst{$^{1}$National Institute of Advanced Industrial Science and Technology, Tsukuba, Ibaraki 305-8568, Japan\\
$^{2}$Faculty of Integrated Arts and Sciences, Tokushima University, Tokushima 770-8502, Japan\\
$^{3}$Institute for Solid State Physics, The University of Tokyo, 106-1 Shirakata, Tokai, Ibaraki 319-1106, Japan\\
$^{4}$Graduate School of Science, Tokyo Metropolitan University, Hachioji, Tokyo 192-0397, Japan}

\abst{The phonon dynamics of filled skutterudite CeRu$_4$Sb$_{12}$ have been studied at room temperature by inelastic neutron scattering. Optical phonons associated with a large vibration of Ce atoms are observed at a relatively low energy of $E \sim$ 6 meV, and show anticrossing behavior with acoustic phonons.  We propose that the origin of the low lattice thermal conductivity in filled skutterudites can be attributed to intensive Umklapp scattering originating from low-lying optical phonons.  By an analysis based on a Born-von K\'{a}rm\'{a}n force model, the longitudinal force constants of the nearest Ce-Sb and Ce-Ru pairs are estimated to be 0.025 mdyn/\AA, while that of the nearest Ru-Sb pair is estimated to be 1.4 mdyn/\AA, indicating that the Ce atoms are bound very weakly to the surrounding rigid RuSb$_6$-octahedron cages.}

\kword{filled skutterudite, rattling motion, phonon, inelastic neutron scattering}

\begin{document}
\maketitle


The rattling phenomenon, that is, \textit{a large vibration of an atom in an oversized atomic cage}, is a key issue in understanding exotic physical properties in a family of bulk materials that possess cage-like units in their crystalline structure.  Filled skutterudites, RM$_4$X$_{12}$ (R = rare-earth; M = Fe, Ru or Os; X = P, As or Sb), are one of those compounds that have large X$_{12}$-icosahedron atomic cages filled with rare-earth atoms \cite{Jeitschko77,Jeitschko80,Lee2004,Lee2004b}.  Despite their well-defined crystalline structures, their lattice thermal conductivity is remarkably small, being comparable to that of vitreous silica \cite{Sales96}.  It has been pointed out by structural refinement studies that the thermal parameters of the filling rare-earth atoms are unusually large, indicating a large vibrational amplitude of the rare-earth atoms \cite{Sales97}.  This result has led to the speculation that the rattling motion of the rare-earth atoms may strongly scatter acoustic phonons, which carry most of the heat flow in a crystal, resulting in an anomalous suppression of the lattice thermal conductivity.

The rattling motion can also influence the electronic properties via electron-phonon coupling.  The filled skutterudites show various interesting features such as heavy fermion superconductivity \cite{Bauer} and a metal-insulator transition \cite{Sekine97}, which could be affected by the rattling.  There are also theoretical works that indicate a correlation between electronic properties and the motion of filled rare-earth atoms \cite{Hattori}.  Recently, it has been proposed that rattling may assist the appearance of superconductivity in the $\beta$-pyrochlore compound KOs$_2$O$_6$ \cite{Yonezawa}.  Charge fluctuation enhanced by rattling is considered to be a possible exotic pairing mechanism responsible for the superconductivity.

Although the importance of rattling phenomena in skutterdites is recognized, there have been only a few studies of rattling motion \cite{Keppens,Hermann,Goto,Kondo,Cao,Iwasa}.  Powder neutron scattering and heat capacity measurements suggest that the rattling motion is an incoherent localized mode, which can be well described by a localized Einstein mode \cite{Keppens,Hermann}.  The energy of the vibrational rattling motion has been estimated from the peak in the phonon density of states measured by powder neutron scattering to be $E = 5 \sim 7$ meV.  Ultrasonic measurements of PrOs$_4$Sb$_{12}$ suggest that a Pr site splits into four off-centered positions, and the Pr atoms vibrate among those positions \cite{Goto}.

To clarify the influence of rattling motion on electronic and thermal transport properties, further studies are quite important.  Inelastic neutron scattering using a single crystal is a powerful method of studying rattling since it can determine both the energy and momentum dependences of phonon spectra.  In this work, we report an inelastic neutron scattering study that uses single crystals of CeRu$_4$Sb$_{12}$ and demonstrate the interplay between acoustic and low-lying optical phonon modes characterized by the large vibration of rare-earth atoms.  We discuss the environment of the filled rare-earth atoms based on the analysis using the Born-von K\'{a}rm\'{a}n method and a possible mechanism of extremely low lattice thermal conductivity in filled skutterudites.



Single crystals of CeRu$_4$Sb$_{12}$ were grown by an Sb flux method, as described elsewhere \cite{Sugawara}.  The size of a single crystal is about 2 $\sim$ 3 mm in length with an almost cubical shape.  To increase the signal intensity in the neutron scattering measurements, six single crystals with a total volume of $\sim 0.2$ cc were assembled.


Neutron scattering measurements were carried out using the triple-axis spectrometer, TOPAN, at the JRR-3M reactor of JAEA at Tokai.  The final neutron energy was fixed at $E_f = 30.5$ meV using a pyrolytic graphite monochromator and an analyzer.  A pyrolytic graphite filter was used to reduce neutrons from higher-order reflections.  The sequences of the horizontal collimators were 15'-15'-S-15'-30' or 40'-60'-S-60'-80', where S denotes the sample position.  The measurements were conducted at room temperature.


For low-energy phonons at below $E = 10$ meV, the observed phonon peaks were fitted using the following scattering function convoluted with the resolution function,
\begin{equation}
S({\bf q},E)=\frac{A}{1-\exp \left(-\frac{E}{k_BT} \right)}
\left\{\frac{\Gamma}{(E-E_s)^2+\Gamma^2} \right\}
\end{equation}
where $E_s$, $\Gamma$, $k_B$, $A$ and $\bf q$ denote the phonon energy, the linewidth, the Boltzmann constant, a scaling factor, and the wave vector, respectively.  In the fitting, we assume that $E_s$ has a linear dependence on $|{\bf q}|$ over the range of the instrumental resolution.  The dynamical structure factor $F_{\mathrm{inel}}$ is estimated using the obtained scattering function.  $F_{\mathrm{inel}}$ is described as
\begin{equation}
F_{\mathrm{inel}} = \sum_{d} \frac{b_d}{\sqrt{m_d}} \exp (-W_d + i {\bf G} \cdot {\bf r_d}) ({\bf Q} \cdot {\bf e_d}),
\end{equation}
where $b_d$ is the coherent scattering length of the $d$th atom at $\bf{r}_d$, $m_d$ is the mass, $e^{-W_d}$ is the Debye-Waller factor, $\bf Q$ is the scattering vector, $\bf G$ is the reciprocal lattice vector, and $\bf{e}_d$ is the polarization vector.  It is known that $F_{\mathrm{inel}}$ is related to the energy-integrated scattering function for a one-phonon process in the neutron energy-loss mode as
\begin{equation}
\int S({\bf Q},E) dE \propto \frac{1}{E_s}
\frac{1}{1-\exp \left(-\frac{E_s}{k_BT} \right)}
|F_{\mathrm{inel}}|^{2}.
\end{equation}

To estimate the interatomic force constants, we have performed calculations based on the Born-von K\'{a}rm\'{a}n atomic force model.  The longitudinal force constants of the seventeen closest atomic pairs were chosen as fitting parameters (see Table \ref{exp-list}), and the calculated intensities of the phonon spectra as well as the energies were fitted to the measured data.  The atomic coordinates of the Sb atoms at 24$g$ sites in the space group $Im\bar{3}$ were assumed to be (0,0.34105,0.15744) \cite{Skutterudite} for the analysis.

\begin{table}[force constant]
\caption{Interatomic force constants obtained from analysis based on Born-von K\'{a}rm\'{a}n model.}
\begin{center}
\begin{tabular}{c c c c}\hline \hline
      &   Pair &   bond length &  Force constants \\
      &          &   (\AA) &  (mdyn/\AA)  \\   \hline
 1 & Ru-Sb   &  2.61 &  1.40  \\
 2 & Sb-Sb   &  2.92 &  0.35  \\
 3 & Sb-Sb   &  2.95 &  0.30  \\
 4 & Ce-Sb   &  3.48 &  0.025 \\
 5 & Sb-Sb   &  3.50 &  0.30  \\
 6 & Sb-Sb   &  3.87 &  0.28  \\
 7 & Ce-Ru   &  4.01 &  0.025  \\
 8 & Sb-Sb   &  4.15 &  0.05  \\
 9 & Ru-Sb   &  4.51 &  0.05  \\
10 &  Ru-Sb  &  4.52 &  0.05  \\
11 &  Sb-Sb  &  4.56 &  0.05  \\
12 &  Ru-Ru  &  4.63 &  0.00  \\
13 &  Sb-Sb  &  5.22 &  0.00  \\
14 &  Sb-Sb  &  5.78 &  0.00  \\
15 &  Ce-Sb  &  5.81 &  0.00  \\
16 &  Sb-Sb  &  5.81 &  0.00  \\
17 &  Ru-Sb  &  5.83 &  0.05  \\ \hline \hline
\end{tabular}
\end{center}
\label{exp-list}
\end{table}%



Figure 1 shows energy spectra observed along $\bf{q}$ $= (\zeta, -\zeta, 0)$ below $E = 10$ meV, which give the transverse phonon modes with the propagation vector $[110]$.  The solid lines depict the calculated profiles from the fits convoluted with the instrumental resolution.  We find that the linewidths of these spectra are comparable to the instrumental resolution.  At $\zeta = 0.15$, a well-defined single peak is observed, which can be assigned to a transverse acoustic (TA) phonon.  Around $\zeta = 0.3$, on the other hand, two peaks are observed with a separation of $\sim1.5$ meV.  The intensity of the lower acoustic mode is strong for small q.  As $\zeta$ increases, however, it gradually decreases and vanishes near the zone boundary by transferring the spectral weight to the higher branch.

\begin{figure}[htb]
\includegraphics[width=\columnwidth]{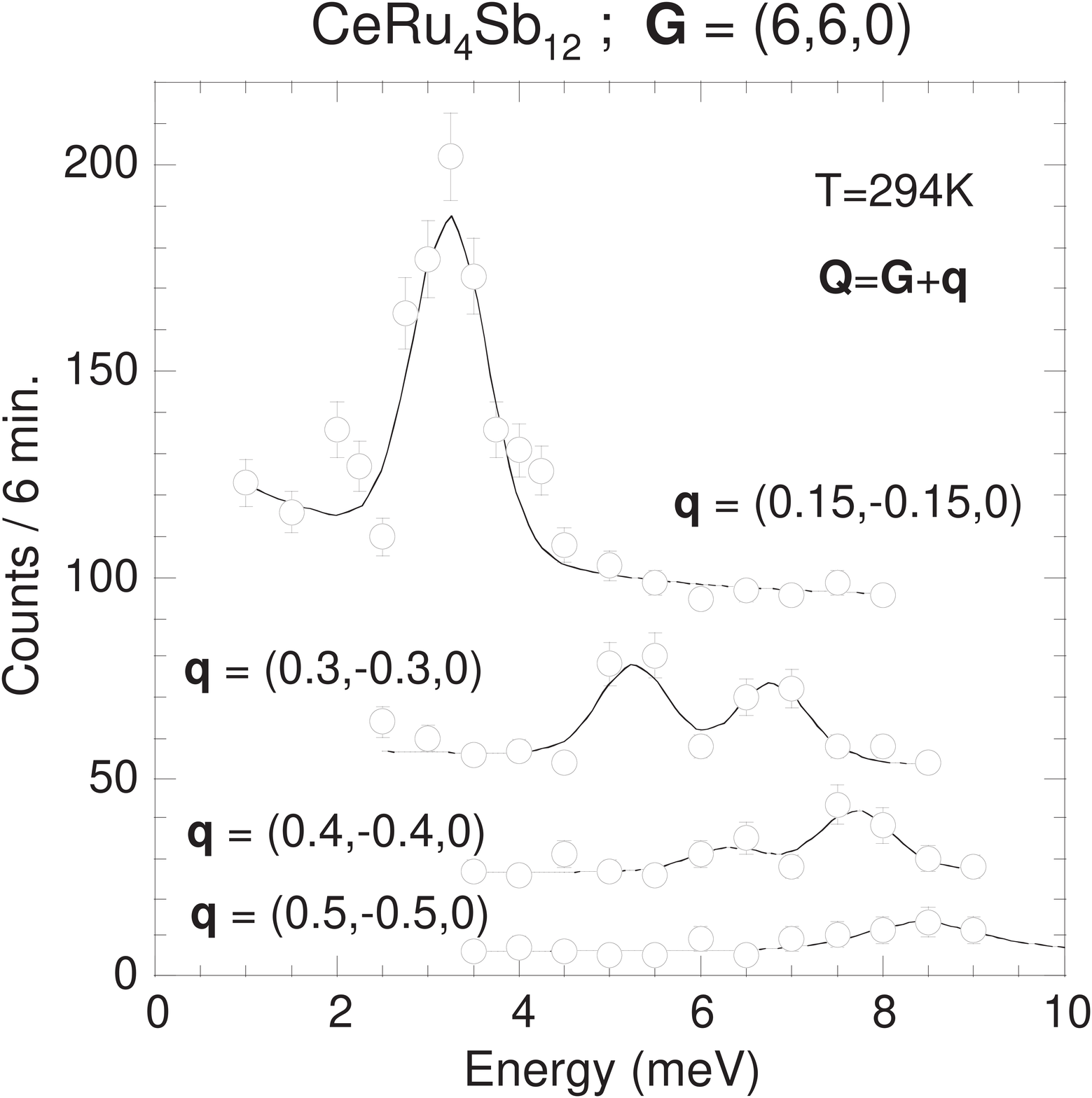}
\caption{\label{fig:acoustic phonons} Energy spectra of transverse acoustic and optical phonon peaks with propagation vector [110].  The solid lines are the results of fits convoluted with the instrumental resolution function.}
\end{figure}

\begin{figure}[htb]
\includegraphics[width=\columnwidth]{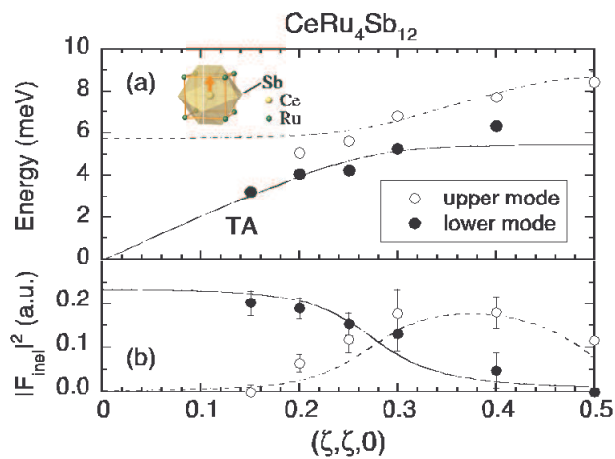}
\caption{\label{fig:low energy dispersion} (a) Phonon dispersion curves of transverse acoustic and optical phonon modes with propagation vector [110] in CeRu$_4$Sb$_{12}$.  The motion of rare-earth ions in the optical mode (guest mode) is indicated by a red arrow in the inset.  (b) Dynamical structure factor as function of $\zeta$.  The solid and dashed lines in (a) and (b) depict the results of a fit based on the Born-von K\'{a}rm\'{a}n model.}
\end{figure}

The dispersion relations of the peaks observed in Fig. \ref{fig:acoustic phonons} are summarized in Fig. \ref{fig:low energy dispersion}(a).  As indicated, the TA and optical modes show typical anticrossing behavior at $\zeta \sim 0.25$. The energy of the lower phonon mode increases linearly with increasing $\zeta$ from the $\Gamma$-point like a typical TA mode.  When the upper phonon mode appears, however, the linear relationship breaks down.  In contrast, the energy of the higher mode increases above $\zeta = 0.2$ and begins to saturate near the zone boundary.  The behavior of the dynamical structure factor shown in Fig. \ref{fig:low energy dispersion}(b) is also consistent with the mixing of two modes. In the region above $\zeta = 0.2$, the dynamical structure factor of the lower energy mode is strongly suppressed, while that of the higher energy mode is enhanced, indicating that they satisfy a sum rule.

Figure 3 shows typical profiles of the transverse optical (TO) modes for energies greater than $E = 10$ meV observed at three $\bf{ q}$-positions with the propagation vector along the [110] direction. Two peaks can be recognized at $E \sim 18$ meV and 35 meV, indicating that the optical phonon modes over $E = 10$ meV are classified into two bands.   The solid lines depict the results of Gaussian fits.  Despite the instrumental energy resolution of $E_{\mathrm{res}} \sim 6$ meV, the full width at half maximum (FWHM) of the phonon profiles at $E \sim 18$ meV is about $\Delta E = 15$ meV, clearly larger than $E_{\mathrm{res}}$, suggesting strongly that the profiles actually consist of a number of closely lying optical branches.

\begin{figure}[htb]
\includegraphics[width=\columnwidth]{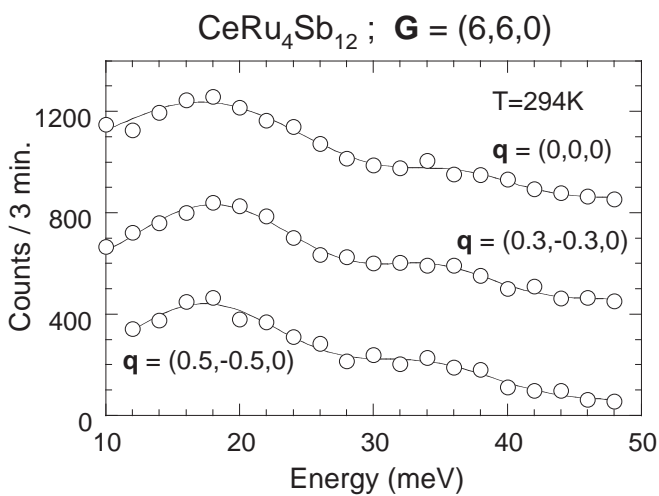}
\caption{\label{fig:optical phonons} Energy spectra of transverse optical phonons with propagation vector along the [110] direction.  The solid lines are the results of Gaussian fits.}
\end{figure}

\begin{figure}[htb]
\includegraphics[width=\columnwidth]{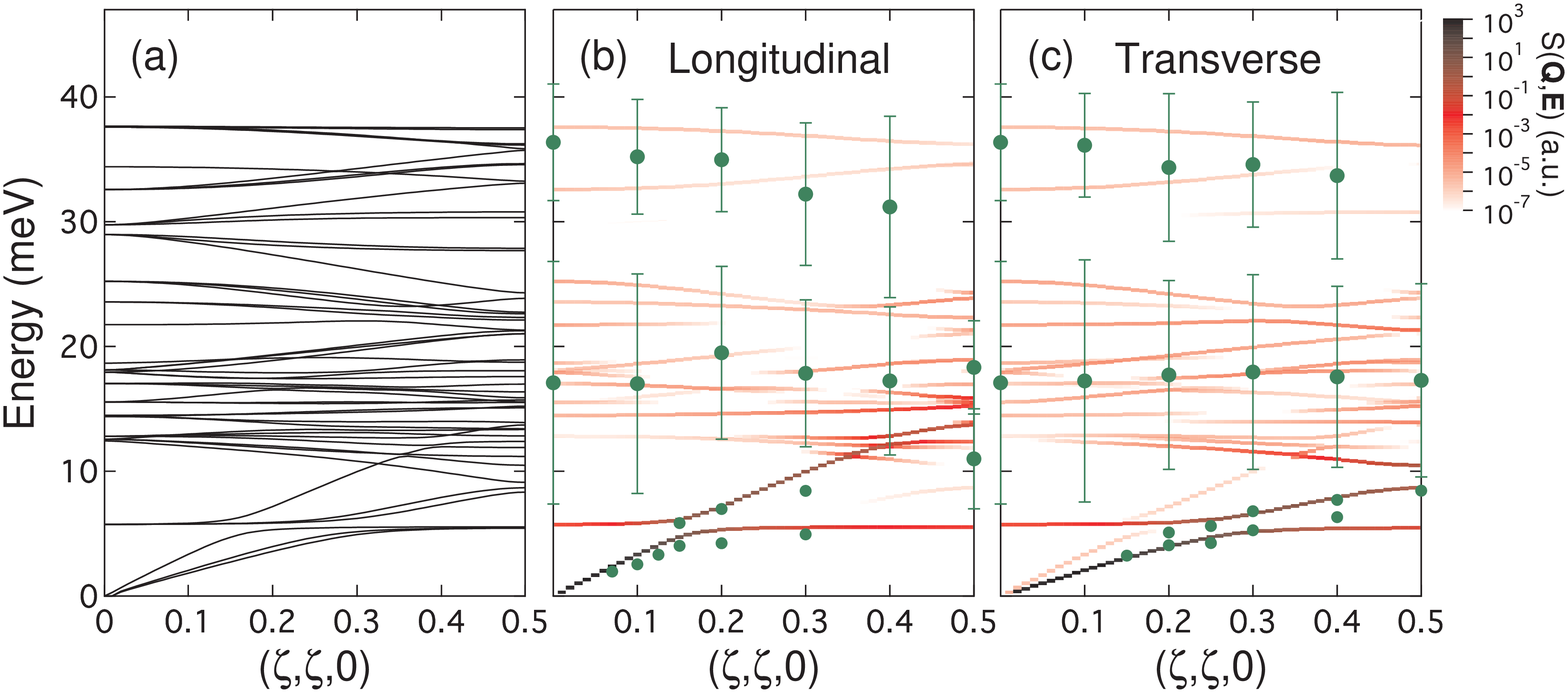}
\caption{\label{fig:dispersion} Phonon dispersion curves along [110] direction in CeRu$_4$Sb$_{12}$.  (a) Results of fit based on the Born-von K\'{a}rm\'{a}n model.  (b) Longitudinal and (c) transverse phonon dispersion curves.  The solid circles depict the results of measurements.  The vertical lines drawn at each data point above $E$ = 10 meV depict the FWHM of each peak.  The contour maps show the calculated scattering function.}
\end{figure}

To interpret the phonon profiles in Figs. \ref{fig:acoustic phonons} and \ref{fig:optical phonons}, we performed analysis based on the Born-von K\'{a}rm\'{a}n model.  From fits to the data, we obtained the calculated phonon dispersion relations.  Those with the propagation vector [110] are depicted in Figs. \ref{fig:low energy dispersion}(a) and \ref{fig:dispersion}(a).  In Figs. \ref{fig:dispersion}(b) and \ref{fig:dispersion}(c), we separately show the dispersion of the longitudinal and transverse modes.  The shade scale of the colored curves corresponds to the spectral weight of the scattering function.  We also depict the peak positions of the observed phonon profiles using green circles.  The vertical lines drawn at each data point above $E = 10$ meV correspond to the FWHM of each phonon profile.  Clearly, the calculated lines follow the observed phonon peaks quite well.  In particular, the anticrossing behavior of the observed phonon dispersions and dynamical structure factor shown in Fig. \ref{fig:low energy dispersion} are successfully reproduced by our Born-von K\'{a}rm\'{a}n force model analysis.  We have confirmed that these anticrossing behaviors can be reproduced only with small Ce-Sb and Ce-Ru force constants.  The alternative lowering of any other force constant leads to large separation between calculation and observation.  The results indicate that the optical mode in the low-energy region at $E \sim 6$ meV is a guest mode with $T_u$ symmetry.  As illustrated in the inset of Fig. \ref{fig:low energy dispersion}(a), the guest mode allows large vibration of the rare-earth atoms.  Since the flat optical band produces a large phonon density of states in the energy spectrum, \textit{the guest mode} can be the vibrational mode observed by inelastic powder neutron scattering measurements at $E = 5 \sim 7$ meV \cite{Keppens,Hermann}, which was identified as being the rattling motion.  Table \ref{exp-list} summarizes the obtained force constants.  The evaluated force constants for Ce-Sb and Ce-Ru pairs are very small, 0.025 mdyn/\AA, reflecting the low phonon energy, and supporting the picture that loosely bound Ce ions rattle within the cage.  However, the anticrossing behavior indicates clearly that the guest atoms interact with host lattices, although the force constants are very small.  This indicates that the vibrations of the guest atoms can propagate to the next guest atoms through host lattices.  The dispersive phonon modes indicate that the guest modes at $E \sim 6$ meV can be interpreted as coherent optical phonon modes rather than a local incoherent Einstein mode.  In the intermediate energy group, the optical phonon energy mainly depends on Sb-Sb force constants.  These force constants have medium values.  In the highest energy group, the optical phonon energy is determined by the large force constants of Ru-Sb pairs, 1.4 mdyn/\AA, indicating that the RuSb$_6$ octahedron is quite rigid.

It has been generally believed that the extremely low lattice thermal conductivity in filled skutterudites is related to the incoherent rattling motions.  Rattled rare-earth atoms were considered to act as scattering centers for acoustic phonons.  This scattering process becomes a dominant mechanism when the rattlers are disordered and/or vibrate incoherently.  However, the Ce atoms in CeRu$_4$Sb$_{12}$ are ordered and fully occupy the Sb$_{12}$-icosahedron atomic cages.  Furthermore, our results suggest that the guest mode is the coherent optical phonon branch, and consequently the Ce atoms cannot function as scattering centers.  In fact, the lattice thermal conductivity of the 100\%-filled samples is higher than that of partially filled samples where rare-earth atoms can scatter acoustic phonons due to disorder \cite{Nolas}.  Feldman \textit{et al}. have also pointed out that the scattering mechanism is not valid for filled skutterudites \cite{Feldman}.  Clearly, another mechanism is needed to explain the low lattice thermal conductivity of 100\%-filled skutterudites.

\begin{figure}[htb]
\includegraphics[width=\columnwidth]{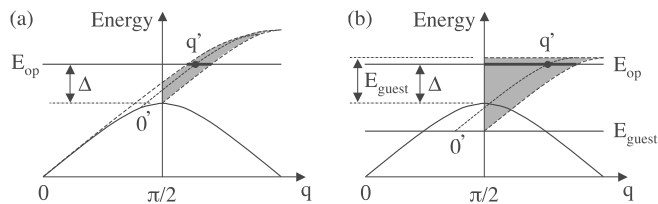}
\caption{\label{fig:Umklapp} Conceptual diagram of phonon dispersion relations.  (a) Simple system without guest modes.  (b) Filled skutterudites with guest modes lying within acoustic phonon branches.  The gray areas indicate the range where a phonon can be created beyond the Brillouin zone via (a) acoustic-acoustic and (b) acoustic-optical phonon scattering.  The thick segments of the flat lines indicate the $\bf{q}$-range of the resultant phonons.  $\Delta$ indicates the gap energy between an acoustic phonon and an upper-lying optical mode.}
\end{figure}

One of the most plausible scattering mechanisms in filled skutterudites is their unique Umklapp processes.  Generally, the phonon modes that contribute to the Umklapp processes are restricted to a narrow $\bf q$-range.  Figure \ref{fig:Umklapp}(a) illustrates a conceptual diagram of phonon dispersion relations in a conventional system.  The gray area depicts a trace of the parallel translation of the original acoustic branch by shifting its origin (O') along the dispersion relation from the zone center towards the zone boundary.  The point $\bf {q'}$ depicts the intersection between the optical and shifted acoustic branches, that satisfy the conservation law. Clearly, the transition of the ``acoustic + acoustic $\rightarrow$ optical phonons'' through the Umklapp process is limited to the short thick line.  In contrast, for filled skutterudites, the flat optical phonon branch lies in a low-energy region below the top of the acoustic phonon branch (Fig. \ref{fig:Umklapp}(b)). Consequently, the process of the ``acoustic + optical $\rightarrow$ optical phonon'' transition is allowed in a wide area, which can be defined by moving the point O' along the low-lying optical phonon branch.  The optical phonons created through this process are distributed along the long $\bf{q'}$ thick line, suggesting that the Umklapp process  occurs more frequently than in Fig. \ref{fig:Umklapp}(a).  It should be noted that this process requires a specific condition: namely, the energy of the guest modes ($E_{\textrm{guest}}$) has to be larger than the gap energy between the acoustic and upper-lying optical phonon modes ($\Delta$).  In fact, $\Delta$ for TA phonons in CeRu$_4$Sb$_{12}$ is $\sim$ 4 meV, which is smaller than $E_{\textrm{guest}}$ $\sim$ 6 meV.  As shown in Fig. \ref{fig:dispersion}(b), this condition $E_{\textrm{guest}} \geq \Delta$ is also satisfied for the longitudinal acoustic mode.  At the same time, a small value of $E_{\textrm{guest}}$ is also essential since the number of phonons contributing to the process increases with decreasing phonon energy.  Since these conditions are all satisfied in the present system, we propose that Umklapp scattering can be one of important processes in the suppression of thermal conductivity in filled skutterudites.   

In conclusion, we have clarified that Ce atoms are weakly bound to surrounding atoms, which ensures their large amplitude of vibration.  The observed phonons at $E \sim$ 6 meV are identified as the optical phonons of a guest mode.  The results suggest that the remarkably low lattice thermal conductivity in filled skutterudites cannot be due to an Einstein oscillation.  As one possibility, we propose that it can be attributed to intensive Umklapp scattering caused by their unique phonon dispersion relations.

The authors would like to thank M. Udagawa, M. Kataoka, Y. Tsunoda and M. Matsuda for their helpful discussions.  This work was supported by a Grant-in-Aid for Scientific Research in Priority Area ``Skutterudite" (Nos. 15072201 and 15072206) of the Ministry of Education, Culture, Sports, Science and Technology of Japan and a grant from the Ministry of Economy, Trade and Industry of Japan.


\begin{thebibliography}{99}

\bibitem{Jeitschko77} W. Jeitschko and D. Braun: Acta Crystallogr., Sect. B: Struct. Crystallogr. Cryst. Chem. \textbf{B33} (1977) 3401.

\bibitem{Jeitschko80} W. Jeitschko and D. Braun: J. Less-Common Met. \textbf{72} (1980) 147.

\bibitem{Lee2004} C. H. Lee, H. Matsuhata, H. Yamaguchi, C. Sekine, and I. Shirotani: J. Magn. Magn. Mater. \textbf{272-276} (2004) 426.

\bibitem{Lee2004b} C. H. Lee, H. Matsuhata, H. Yamaguchi, C. Sekine, K. Kihou, T. Suzuki, T. Noro, and I. Shirotani: Phys. Rev. B \textbf{70} (2004) 153105.

\bibitem{Sales96} B. C. Sales, D. Mandrus, and R. K. Williams: Science \textbf{272} (1996) 1325.

\bibitem{Sales97} B. C. Sales, D. Mandrus, B. C. Chakoumakos, V. Keppens, and J. R. Thompson: Phys. Rev. B \textbf{56} (1997) 15081.

\bibitem{Bauer} E. D. Bauer, N. A. Frederick, P. C. Ho, V. S. Zapf, and M. B. Maple: Phys Rev. B \textbf{65} (2002) 100506.

\bibitem{Sekine97} C. Sekine, T. Uchiumi, I. Shirotani, and T. Yagi: Phys. Rev. Lett. \textbf{79} (1997) 3218.

\bibitem{Hattori} K. Hattori and K. Miyake: J. Phys. Soc. Jpn. Suppl. \textbf{75} (2006) 238.

\bibitem{Yonezawa} S. Yonezawa, Y. Muraoka, Y. Matsushita, and Z. Hiroi: J. Phys.: Condens. Matter \textbf{16} (2004) L9.

\bibitem{Keppens} V. Keppens, D. Mandrus, B. C. Sales, B. C. Chakoumakos, P. Dai, R. Coldea, M. B. Maple, D. A. Gajewski, E. J. Freeman, and S. Bennington: Nature (London) \textbf{395} (1998) 876.

\bibitem{Hermann} R. P. Hermann, R. Jin, W. Schweika, F. Grandjean, D. Mandrus, B. C. Sales, and G. J. Long: Phys. Rev. Lett. \textbf{90} (2003) 135505.

\bibitem{Goto} T. Goto, Y. Nemoto, K. Sakai, T. Yamaguchi, M. Akatsu, T. Yanagisawa, H. Hazama, K. Onuki, H. Sugawara, and H. Sato: Phys. Rev. B \textbf{69} (2004) 180511.

\bibitem{Kondo} T. Kondo, K. Yamamoto, N. Ogita, M. Udagawa, H. Sugawara, and H. Sato: Physica B \textbf{359-361} (2005) 904.

\bibitem{Cao} D. Cao, F. Bridges, P. Chesler, S. Bushart, E. D. Bauer, and M. B. Maple: Phys. Rev. B \textbf{70} (2004) 94109.

\bibitem{Iwasa} K. Iwasa, M. Kohgi, H. Sugawara, and H. Sato: Physica B \textbf{378-380} (2006) 194.

\bibitem{Sugawara} H. Sugawara, S. Osaki, S. R. Saha, Y. Aoki, H. Sato, Y. Inada, H. Shishido, R. Settai, Y. Onuki, H. Harima, and K. Oikawa: Phys. Rev. B \textbf{66} (2002) 220504.

\bibitem{Skutterudite} K. Oikawa: {\it Skutterudite News Letter}, eds. H. Sato and H. Harima: Vol. \textbf{2} (2004) No. 2, p. 8.

\bibitem{Nolas} G. S. Nolas, J. L. Cohn, and G. A. Slack: Phys. Rev. B \textbf{58} (1998) 164.

\bibitem{Feldman} J. L. Feldman, D. J. Singh, I. I. Mazin, D. Mandrus, and B. C. Sales: Phys. Rev. B \textbf{61} (2000) 9209.

\end{thebibliography}
\end{document}